\documentclass{WileyMSP-template}
\usepackage{amsmath,amssymb,amsthm}
\usepackage{graphicx}
\usepackage[utf8]{inputenc}
\usepackage[english]{babel}
\usepackage{xcolor} 
\makeatletter
\let\latexps@plain\ps@plain
\newcommand{\frontmatter}{\let\ps@plain\ps@empty\pagestyle{empty}}
\newcommand{\mainmatter}{%
  \let\ps@plain\latexps@plain\pagestyle{plain}%
  \clearpage
  \pagenumbering{arabic}}
\makeatother
\begin{document}

\pagestyle{fancy}
\rhead{\includegraphics[width=2.5cm]{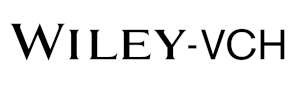}}

\title{Understanding infection progression under strong control \\ measures through universal COVID-19 growth signatures}

\maketitle

\author{Magdalena Djordjevic*}
\author{Marko Djordjevic*}
\author{Bojana Ilic}
\author{Stefan Stojku}
\author{Igor Salom}



\begin{affiliations}
Dr. Magdalena Djordjevic\\
Institute of Physics Belgrade, University of Belgrade, Serbia\\
Email Address: magda@ipb.ac.rs

Prof. Marko Djordjevic\\
Quantitative Biology Group, Faculty of Biology, University of Belgrade, Serbia\\
Email Address: dmarko@bio.bg.ac.rs

Dr. Bojana Ilic\\
Institute of Physics Belgrade, University of Belgrade, Serbia\\

Stefan Stojku\\
Institute of Physics Belgrade, University of Belgrade, Serbia\\

Dr. Igor Salom\\
Institute of Physics Belgrade, University of Belgrade, Serbia\\

\end{affiliations}


\keywords{Infections disease modeling; Dynamical growth patterns; Physics and society; Scaling of epidemics growth}

\begin{abstract}

Widespread growth signatures in COVID-19 confirmed case counts are reported, with sharp transitions between three distinct dynamical regimes (exponential, superlinear and sublinear). Through analytical and numerical analysis, a novel framework is developed that exploits information in these signatures. An approach well known to physics is applied, where one looks for common dynamical features, independently from differences in other factors. These features and associated scaling laws are used as a powerful tool to pinpoint regions where analytical derivations are effective, get an insight into qualitative changes of the disease progression, and infer the key infection parameters. The developed framework for joint analytical and numerical analysis of empirically observed COVID-19 growth patterns can lead to a fundamental understanding of infection progression under strong control measures, applicable to outbursts of both COVID-19 and other infectious diseases.

\end{abstract}


\textbf{\textit{Introduction: }}
COVID-19 pandemic introduced unprecedented worldwide social distancing measures\textsuperscript{\cite{1}}. While interventions such as quarantine or vaccination have been extensively studied in quantitative epidemiology, effects of social distancing are not well understood\textsuperscript{\cite{2}-\cite{4}}, and when addressed, they have been studied only numerically.  Unique opportunity to understand these effects has been provided by COVID-19 tracing through confirmed case counts, active cases and fatalities, in a variety of countries with different demographic and environmental conditions\textsuperscript{\cite{5,6}}. We here show that focusing on analytical and numerical derivations in distinct epidemics growth regimes, is a novel and effective approach in revealing infection progression mechanisms
that may be a valuable alternative to detailed numerical simulations.

\textbf{\textit{Outline: }}
Our COVID-19 dynamics model is introduced in the Model section.  In Numerical Framework/Results and Discussion sections, we will extract COVID-19 count data~\cite{Worldometer} and select those countries that systematically trace not only confirmed cases and fatalities, but also active cases (Andorra, Austria, Czechia, Croatia, Cuba, Germany, Israel, New Zealand, Switzerland and Turkey), which allows tight constraint of numerical analysis. We will observe three characteristic growth regimes in confirmed case counts, show that our model is well constrained by these regimes for a wide range of countries, and provide an intuitive explanation behind the emergence of such regimes. In Analytical Framework/Results and Discussion sections, analysis of the characteristic (inflection and maximum) points of the infective curve will allow to i) explain the nearly constant value of the scaling exponent in the superlinear regime of confirmed counts; {\it ii)} understand the relation between the duration of this regime and strength of social distancing; {\it iii)} pinpoint changes in the reproduction number from outburst to extinguishing the infection, and {\it iv)} constrain the main parameter quantifying the effect of social distancing by analysing scaling of the infection growth with time in the sublinear regime.  The obtained constraints provide a basis for successful analysis of countries that did not continuously track the active cases (here demonstrated for France,
Italy, Spain, United Kingdom and Serbia). We will finally present the key infection parameters inferred through combined analytical and numerical analysis.

\textbf{\textit{Model: }}
We develop a mechanistic model (nonlinear and nonhomogeneous), which takes into account gradual introduction of social distancing (as relevant for most countries' response), in addition to other important infection progression mechanisms. We start from standard compartments for epidemiological models, i.e., susceptible ($S$), exposed ($E$), infective ($I$) and recovered ($R$)\textsuperscript{\cite{2}-\cite{4}}. To account for social distancing and observable quantities, we  introduce additional compartments: protected ($P$) - where individuals effectively move from susceptible category due to social distancing; total number of diagnosed (confirmed and consequently quarantined) cases ($D$), active cases ($A$),  and fatalities ($F$). $D$, $A$ and $F$  correspond to directly observable (measured) quantities, but are indirect observables of $I$, as only part of infective individuals gets diagnosed, due to a large number of mild/asymptomatic cases\textsuperscript{\cite{7}}.

 We implement the model deterministically, as COVID-19 count numbers are very high wherever reasonable testing capacities are employed. This makes model analytically tractable, and allows robust parameter inference through combination of analytically derived expressions and tightly constrained numerical analysis, as we show below. Our analysis is applied separately to each country, as the effect of social distancing, initial numbers of infected and exposed cases, diagnosis/detection efficiency and transmission rates may be different. However, within a given country, we do not take into account different heterogeneities $-$ demographic, spatial, population activity, or seasonality effects\textsuperscript{\cite{2,8,9}}. Alternatively, global dynamical properties of the outbreak can be analyzed in a probabilistic framework employing partial differential equations in an age-structured model\textsuperscript{\cite{24,25}}. These can readily be included in our model,
but would lead to model structure which is not analytically tractable,
so these extensions are left for future work.

Given this, the model equations are:
\begin{align}~\label{jne1}
&dS/dt = - \beta I S/N - dP/dt; \,\,\,\, dP/dt =  \alpha/(1+(t_0/t)^n) S  \\
&dE/dt =\beta I S/N -\sigma E; \,\,\,\,
dI/dt =\sigma E -\gamma I -\epsilon \delta I; \,\,\,\, dR/dt = \gamma I\\
&dD/dt = \epsilon \delta I; \,\,\,\,dA/dt = \epsilon \delta I -h A - m A;
 \,\,\,\, dF/dt = m A
\end{align}
where $N$ is the total population number; $\beta$ - the transmission rate; $\sigma$ - inverse of the latency period; $\gamma$ - inverse of the infectious period; $\delta$ - inverse of the detection/diagnosis period; $\epsilon$ - detection efficiency; $h$ - the recovery rate; $m$ - the mortality rate. Social distancing is included through Eq. (1) (second equation), which represents the rate at which the population moves (on average) from  susceptible to protected category. The term $\frac{ \alpha}{1+(t_0/t)^n}$ corresponds to a sigmoidal dependence (similar to Fermi-Dirac function, in quantitative biology known as the Hill function\textsuperscript{\cite{10}}). Time $t_0$ determines the half-saturation, so that well before $t_0$ the social distancing is negligible, while well after $t_0$ the rate of transition to the protected category approaches $\alpha$. Parameter $n$ (the Hill constant) determines how rapidly the social distancing is introduced, i.e., large $n$ leads to rapid transition from OFF to ON state, and vice versa\textsuperscript{\cite{10}}. Eq. (3) considers that only a fraction of the infected is diagnosed, so that $\epsilon \delta I$ takes into account the diagnosis and the subsequent quarantine process.

\textbf{\textit{Analytical Framework: }}
To make the problem analytically tractable, we approximate the Hill function in the first relation of Eq.~(1) by unit step function, so that after $t_0$ the $2^{nd}$ term in Eq.~(1) becomes $-\alpha S$ and dominates over the $1^{st}$ term, i.e., $S(t)\sim e^{-\alpha t}$. We checked that this approximation agrees well with full-fledged numerical simulations (Figure 1D and Supplement). In all comparisons with analytical results, numerical analysis is done with the full model, allowing an independent check of both analytical derivations and employed approximations. Under this assumption, Eqs. (1-2) reduce to:
\begin{align} \label{SODE}
\frac{d^2 I(t)}{dt^2} + (\gamma+\epsilon \delta + \sigma) \frac{dI(t)}{dt} = \sigma \Big\{ \beta [\theta(t_0-t) + e^{-(t-t_0) \alpha} \theta(t-t_0)] - (\gamma + \epsilon \delta) \Big\} I(t).
\end{align}
We next introduce two time regions: I) $t\leq t_0$ and II) $t>t_0$ and solve Eq.~\eqref{SODE} separately within these regions, where corresponding solutions are denoted as $I_I(t)$ and $I_{II}(t)$. As in the above expressions $\gamma+\epsilon \delta$ always appear together, we further denote $\gamma+\epsilon \delta \rightarrow \gamma$.

For $I_I(t)$, we take $I(t=0)\equiv I_0$, and restrict to dominant (positive) Jacobian eigenvalue, leading to the exponential regime:
\begin{align}~\label{IprviReg}
I_I(t)=I_0 e^{\frac{1}{2}[ -(\gamma + \sigma) +\sqrt{(\gamma - \sigma)^2 + 4 \beta \sigma}] t}.
\end{align}
By shifting $t-t_0 \rightarrow t$, $I_{II}(t)$ is determined by
\begin{align}~\label{SODE2}
\frac{d^2 I_{II}(t)}{dt^2} + (\gamma + \sigma) \frac{dI_{II}(t)}{dt} = \sigma (\beta e^{-\alpha t } - \gamma  ) I_{II}(t).
\end{align}

Eq.~\eqref{SODE2} is highly nontrivial, due to variable coefficient ($\sigma \beta e^{-\alpha t }$). By substituting variable $t\rightarrow x= \frac{-2 i \sqrt{\beta \sigma}}{\alpha} e^{-\frac{\alpha t}{2}}$ it can be shown that Eq.~\eqref{SODE2} reduces to transformed form of Bessel differential equation\textsuperscript{\cite{11}}:
\begin{align}~\label{MBDE}
x^2 \frac{d^2 y}{dx^2} + (1-2 \alpha_1) x \frac{dy}{dx} +  \Big( \beta^2_1 \gamma^2_1 x^{2 \gamma_1} + \alpha^2_1 -\nu^2 \gamma^2_1 \Big) y=0,
\end{align}
whose general solution for noninteger $\nu$ is given by:
\begin{align}~\label{MBDEs}
y(x)=x^{\alpha_1} \Big[C_1 J \Big(\nu,\beta_1 x^{\gamma_1} \Big) +C_2 J \Big(-\nu,\beta_1 x^{\gamma_1} \Big) \Big],
\end{align}
where $J \big(\nu,x \big) $ represents Bessel function of the first kind, and $C_1, C_2$ are arbitrary constants. In our case $\alpha_1=\frac{\gamma +\sigma}{\alpha}$, $\gamma_1=\beta_1=1$, while $\nu=\frac{\gamma -\sigma}{\alpha}$ is indeed nonintiger. If we return to $t$ variable, taking into account the following relation between standard and modified ($I\big(\nu,x \big)$) Bessel functions of the first kind\textsuperscript{\cite{12,13}}: $I\big(\nu,x \big)=i^{-\nu} J\big(\nu,ix \big)$, the general solution of Eq.~\eqref{SODE2} reads:
 \begin{align}~\label{IdrugiReg0}
I_{II}(t)& {}  = \Big(\frac{\beta \sigma}{\alpha^2} e^{-\alpha t} \Big)^{\frac{\gamma +\sigma}{2 \alpha}} \Big\{C_1 (-1)^{\frac{\gamma}{\alpha}} I\Big(\frac{\gamma - \sigma}{\alpha}, \frac{2\sqrt{e^{-\alpha t} \beta \sigma}}{\alpha} \Big) \Gamma \Big(1+\frac{\gamma - \sigma}{\alpha} \Big)\nonumber \\
& + C_2 (-1)^{\frac{\sigma}{\alpha}} I\Big(-\frac{\gamma - \sigma}{\alpha}, \frac{2\sqrt{e^{-\alpha t} \beta \sigma}}{\alpha} \Big) \Gamma \Big(1-\frac{\gamma - \sigma}{\alpha} \Big) \Big\}.
\end{align}
To determine $C_1$, $C_2$, we use the following boundary conditions: $I_{II}(0)=I_I(t_0)$ and $I'_{II}(0)=I'_I(t_0)$, where the first derivative in region II has the following expression:
\begin{align}~\label{DrugiRegionPrviIzvod}
I'_{II}(0) ={} & \Big(\frac{\beta \sigma}{\alpha^2}  \Big)^{\frac{\gamma +\sigma}{2 \alpha}} \Big \{C_1 (-1)^{\frac{\alpha+\gamma}{\alpha}} \Gamma\Big(1+\frac{\gamma - \sigma}{\alpha} \Big) \Big[\gamma I\Big(\frac{\gamma - \sigma}{\alpha}, \frac{2\sqrt{ \beta \sigma}}{\alpha} \Big)
+ \sqrt{\beta \sigma} I\Big(1+\frac{\gamma - \sigma}{\alpha}, \frac{2\sqrt{ \beta \sigma}}{\alpha} \Big) \Big]\nonumber \\
& + C_2 (-1)^{\frac{\alpha+\sigma}{\alpha}}\Gamma\Big(1-\frac{\gamma - \sigma}{\alpha} \Big) \Big[\sigma I\Big(-\frac{\gamma - \sigma}{\alpha}, \frac{2\sqrt{ \beta \sigma}}{\alpha} \Big)
+ \sqrt{\beta \sigma} I\Big(1-\frac{\gamma - \sigma}{\alpha}, \frac{2\sqrt{ \beta \sigma}}{\alpha} \Big) \big] \Big\}.
\end{align}
In obtaining the expression above, the following identities were frequently used\textsuperscript{\cite{12,13}}:
\begin{align}~\label{Balgebra1}
\frac{dI\big(\nu,x\big)}{dx} ={} I\big(\nu-1,x\big) - \frac{\nu}{x} I\big(\nu,x\big); \,\,\,\,
I\big(\nu-1,x\big) - I\big(\nu+1,x\big)  = \frac{2 \nu I\big(\nu,x\big)}{x}.
\end{align}
After derivations, where the following relation\textsuperscript{\cite{13}}
\begin{align}~\label{Balgebra3}
I\big(\nu+1,x\big) I\big(-\nu,x\big)- I\big(\nu,x\big) I\big(-\nu-1,x\big)={} &\frac{2 \sin(\pi \nu)}{\pi x},
\end{align}
together with $\sin((\nu \pm1) \pi)=-\sin(\nu \pi)$ and the identity relating modified Bessel function of the first and second kind
$K\big(\nu,x\big)= \frac{\pi}{2} \frac{I(-\nu,x) -I(\nu,x)}{\sin{\nu \pi}}$
are used\textsuperscript{\cite{12,13}}, we finally obtain a surprisingly simple result:
\begin{align}~\label{IdrugiReg}
I_{II}(t)& {}  = I_I(t_0) e^{-\frac{\gamma + \sigma}{2}t} \frac{ K\Big(\frac{\gamma - \sigma}{\alpha}, \frac{2\sqrt{e^{-\alpha t} \beta \sigma}}{\alpha} \Big)}{K\Big(\frac{\gamma - \sigma}{\alpha},\frac{2\sqrt{\beta \sigma}}{\alpha}\Big)},
\end{align}
where $K\big(\nu,x \big)$ is the modified Bessel function of the $2^{nd}$ kind.

At maximum and inflection points, $I'_{II} =0$ and $I''_{II} =0$, respectively. After extensive simplification of the results, this leads to ($y=R_{0,\rm{free}} e^{-\alpha t}$, where $R_{0,\rm{free}}=\beta/\gamma $ is the basic reproduction number in the absence of social distancing\textsuperscript{\cite{6,17}}):
\begin{align}~\label{jednakost11}
\sqrt{y} K\Big(\frac{\gamma - \sigma}{\alpha} + 1, \frac{2\sqrt{\gamma \sigma}}{\alpha} \sqrt{y} \Big)& {}  =
 \sqrt{\frac{\gamma}{\sigma}} K\Big(\frac{\gamma - \sigma}{\alpha}, \frac{2\sqrt{\gamma \sigma}}{\alpha} \sqrt{y} \Big),
 \end{align}
\begin{align}~\label{jednakost22}
\sqrt{y} K\Big(\frac{\gamma - \sigma}{\alpha} + 1, \frac{2\sqrt{\gamma \sigma}}{\alpha} \sqrt{y} \Big) =  \sqrt{\frac{\gamma}{\sigma}}
\frac{(\frac{\gamma}{\sigma} +y)}{(\frac{\gamma}{\sigma}+1)} K\Big(\frac{\gamma - \sigma}{\alpha}, \frac{2\sqrt{\gamma \sigma}}{\alpha} \sqrt{y} \Big).
\end{align}

Eqs.~(\ref{jednakost11},~\ref{jednakost22}) have to be solved numerically, but, as $\gamma$ and  $\sigma$ are constants, we, interestingly, obtain that solutions will depend only on $\alpha$. Since, for the analysis of superlinear and sublinear regimes, only the left inflection point and the maximum are important, we will further omit the second solution of Eq.~\eqref{jednakost22} (Eq.~\eqref{jednakost11} has one solution), and denote $y_i =f_{i1}(\alpha)\equiv f_i(\alpha)$, $y_m=f_m(\alpha)$ (these two solutions are presented as upper and lower curves on Figure~2C, respectively), so that the effective reproduction numbers at inflection and maximum points ($R_{e, i}$ and $R_{e, m}$) are:
\begin{align}~\label{EFT}
R_{e, i} \equiv {} & R_{0,\rm{free}} e^{-\alpha t_i}  = f_i(\alpha), \nonumber \\
R_{e, m} \equiv {} & R_{0,\rm{free}} e^{-\alpha t_m}  = f_m(\alpha).
\end{align}
From this follows the length of superlinear regime (between inflection and maximum points):
\begin{align}~\label{deltaT}
\Delta t \equiv t_m -t_i = \frac{1}{\alpha} \ln{\Big(\frac{f_i(\alpha)}{f_m(\alpha)} \Big)}.
\end{align}

\begin{figure*}
  \vspace*{-0.8cm}
  \includegraphics[width=\linewidth]{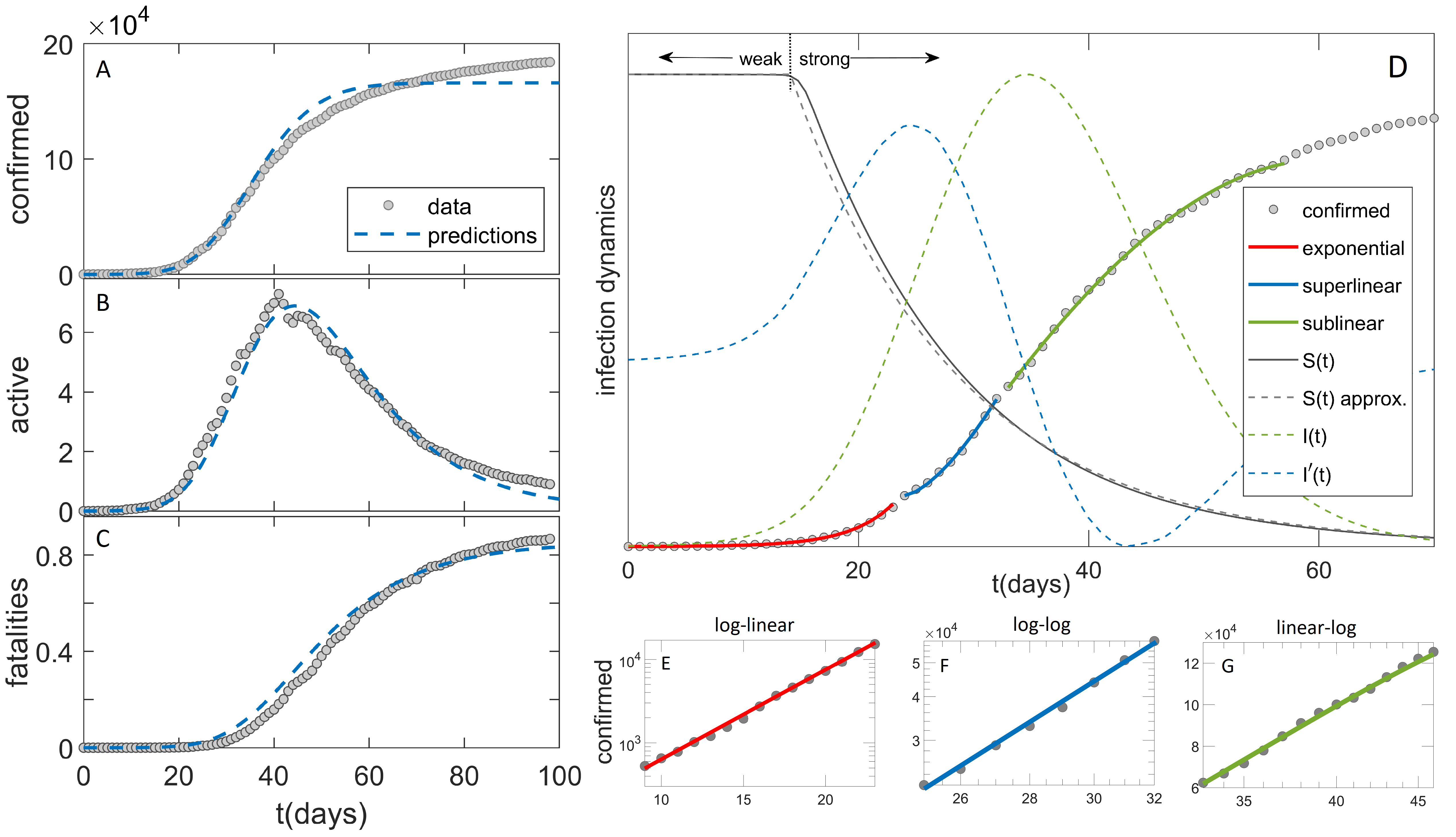}
  \caption{ Comparison of the model (dashed blue curves) with the data in the case of Germany (grey circles) for A) confirmed case counts, B) active cases, C) fatalities. D) Exponential, superlinear and sublinear fit to confirmed case data, is shown. Arrows "weak" and "strong" indicate, respectively, the regions with a small and large magnitude of social distancing. The full grey curve denotes susceptibles ($S(t)$), where the dashed grey curve shows an approximation to $S(t)$. The dashed green curve denotes the number of infectious cases ($I(t)$), where the dashed blue curve is $I'(t)$, whose maxima indicate $I(t)$ inflection points. The confirmed case counts in the three regimes are shown on E) log-linear, F) log-log and G) linear-log scale.}
  \label{fig:sl1}
\end{figure*}
We further Taylor expand $I_{II}(t)$ around the inflection point:
\begin{align}~\label{IdrugiRegn}
I_{II}(t) & {}  = I_I(t_0) e^{-\frac{\gamma + \sigma}{2}t_i} \frac{K\Big(\frac{\gamma - \sigma}{\alpha}, \frac{2 \sqrt{\gamma \sigma}}{\alpha} \sqrt{f_i(\alpha)} \Big)}{K\Big(\frac{\gamma - \sigma}{\alpha}, \frac{2 \sqrt{\beta \sigma}}{\alpha} \Big)}
\Big[1- \frac{\gamma \sigma}{\gamma + \sigma} \Big(1-f_i(\alpha) \Big) (t-t_i) +\mathcal{O}\big((t-t_i)^2\big) \Big] .
\end{align}
In the superlinear regime $D(t)\sim(t-t_s)^\upsilon$, where $\upsilon$ is the scaling exponent and $t_s$ marks the beginning of this regime. By Taylor expanding $D(t)$ around $t_i$, using Eq.~\eqref{IdrugiRegn} and Eq. (3):
\begin{align}~\label{expV}
& \upsilon =  1+ \frac{1}{k \alpha} \frac{\gamma \sigma}{\gamma + \sigma} [f_i(\alpha) -1 ] \ln{\Big( \frac{f_i(\alpha)}{f_m(\alpha)} \Big)},
\end{align}
which is always larger than 1, as expected for the superlinear regime. As $t_i$ is localized towards the beginning of the regime, we estimate $t_i -t_s \sim \frac{\Delta t}{k}$, where $k \approx 3,4$.

Finally, to provide analytical constrain on $\alpha$, we Taylor expand $I_{II}(t)$ around the maximum:
\begin{align}~\label{MP_T}
I_{II}(t)& {}  = I_I(t_0)e^{-\frac{\gamma + \sigma}{2}t_m} \frac{K\Big(\frac{\gamma - \sigma}{\alpha}, \frac{2 \sqrt{\gamma \sigma}}{\alpha} \sqrt{f_m(\alpha)} \Big)}{K\Big(\frac{\gamma - \sigma}{\alpha}, \frac{2 \sqrt{\beta \sigma}}{\alpha} \Big)}
\Big[1- \frac{\gamma \sigma}{2} \Big(1-f_m(\alpha) \Big) (t-t_m)^2 +\mathcal{O}\big((t-t_m)^3\big) \Big] .
\end{align}
As $f_m(\alpha) < 0$, we see that the quadratic term in Eq.~\eqref{MP_T} is always negative, i.e., $D(t)$ curve enters sublinear regime around maximum of the infection.  By fitting $D(t)$ to $c+d(t-t_m) - f(t-t_m)^3$ in this regime, and by using Eq.~\eqref{MP_T} together with Eq.~(3), we obtain:
\begin{align}~\label{jednakost3}
\frac{f}{d} & {} =\frac{\gamma \sigma}{6} [1-f_m(\alpha)],
\end{align}
which allows to directly constrain $\alpha$.

\textbf{\textit{Numerical Framework: }} We first numerically analyze outburst dynamics in the countries that continuously updated\textsuperscript{\cite{14}} {\it three} observable categories ($D$, $A$ and $F$). For a large majority of countries active cases were either not tracked or were not continuously updated, so the analysis is done for 10 countries listed in the Outline section.

In the exponential regime, the analytical closed-form solution is given by Eq.~\eqref{IprviReg}. From this, and the initial slope of $\ln{(D)}$ curve (once the number of counts are out of the stochastic regime), $\beta$ can be directly determined, while the corresponding eigenvector sets the ratio of $I_0$ to $E_0$. The intercept of the initial exponential growth of $D$ at $t=0$ sets the product of $I_0$ and $\epsilon \delta$. $h$ and $m$ can also be readily constrained, as from Eqs.~(3), they depend only on integrals of the corresponding counts; here note that $d(D-A-F)/dt = h A$. Also\textsuperscript{\cite{17},\cite{15},\cite{16}}, $\sigma=1/3$~day$^{-1}$ and $\gamma=1/4$~day$^{-1}$, characterize fundamental infectious process, which we assume not to change between different countries.

Only parameters related with the intervention measures ($\alpha, t_0, n, \epsilon \delta$) are left to be inferred numerically, leading to tightly constrained numerical results. For this, we individually performed joint fit to all three observable quantities ($A, D, F$) for each country. The errors are estimated through Monte-Carlo\textsuperscript{\cite{18,19}} simulations, assuming that count numbers follow Poisson distribution.

\textbf{\textit{Results and Discussion: }}
Representative numerical results are shown in Figure~1 for Germany, while other countries are shown in the Supplement. In Figure~1A-C (and Supplement) we see a good agreement of our numerical analysis with all three classes of the case counts. In Figure~1D, we see sharp transitions between the three growth patterns indicated in the figure: {\it i)} {\it exponential} growth, observed as a straight line in log-linear plot in Figure~1E; {\it ii)} {\it superlinear} growth, a straight line in log-log plot in Figure~1F; {\it iii) sublinear} growth, a straight line in linear-log plot in Figure~1G.

Transition between the growth patterns can be qualitatively understood from Eq. (3), and $I(t)$ curve in Figure~1D. The exponential growth has to break after the inflection point of $I(t)$, i.e., once the maximum of its first derivative ($I'(t)$ in Figure 1D) is reached. In the {\it superlinear regime}, confirmed counts case ($D(t)$) curve is convex ($D''(t) > 0$), so this regime breaks once $I'(t)$ (dashed blue curve) becomes negative. Equivalently, $D(t)$ curve becomes concave (enters {\it sublinear regime}) once the maximum of the $I(t)$ is reached. Note that the growth of $D(t)$ can reemerge if the social distancing measures are alleviated. Our model can account for this by allowing transition from protected back to susceptible category, which is out of the scope of this study, but may improve the agreement with the data at later times (see Figs. 1A-C). In addition to this numerical/intuitive understanding, we also showed that we analytically reproduce the emergence of these growth regimes (Eqs.~(6), (14), (15)). Can we also analytically derive the parameters that characterize these regimes?

The {\it exponential regime} is straightforward to explain, as described in the previous section. The {\it superlinear} regime is in between the left inflection point and the maximum of $I(t)$, so that infective numbers grow, but with a decreasing rate. While the derivations are straightforward in the exponential regime, they are highly non-trivial during the subsequent subexponential (superlinear and sublinear) growth.
As the superlinear regime spans the region between the left inflection point ($t_i, I''(t_i)=0$) and the maximum ($t_m, I'(t_m)=0$), its duration is $\Delta t = t_m -t_i$ given by Eq.~(\ref{deltaT}), with $\sim 1/\alpha$ dependence, so that weak measures lead to protracted superlinear growth (see Figure~2A). This tendency is also confirmed by independent numerical analysis in Figure~2A, where for each individual country we numerically infer $\alpha$ and extract the length of the superlinear regime. Therefore, the duration of the superlinear regime indicates the effectiveness of introduced social distancing.

The scaling exponent $\upsilon$ of the superlinear regime is given by Eq.~(\ref{expV}), and shown in Figure 2B, where we predict that all countries are roughly in the same range of $1.2 < \upsilon < 1.5$ (surprisingly, weakly dependent on $\alpha$), despite significant differences in the applied measures, demographic and environmental factors.  This result is (independently from our model) confirmed from case count numbers (the slope in Figure 1F, and equivalently for other countries, see Figure 2B).

\begin{figure*}
  \includegraphics[width=\linewidth]{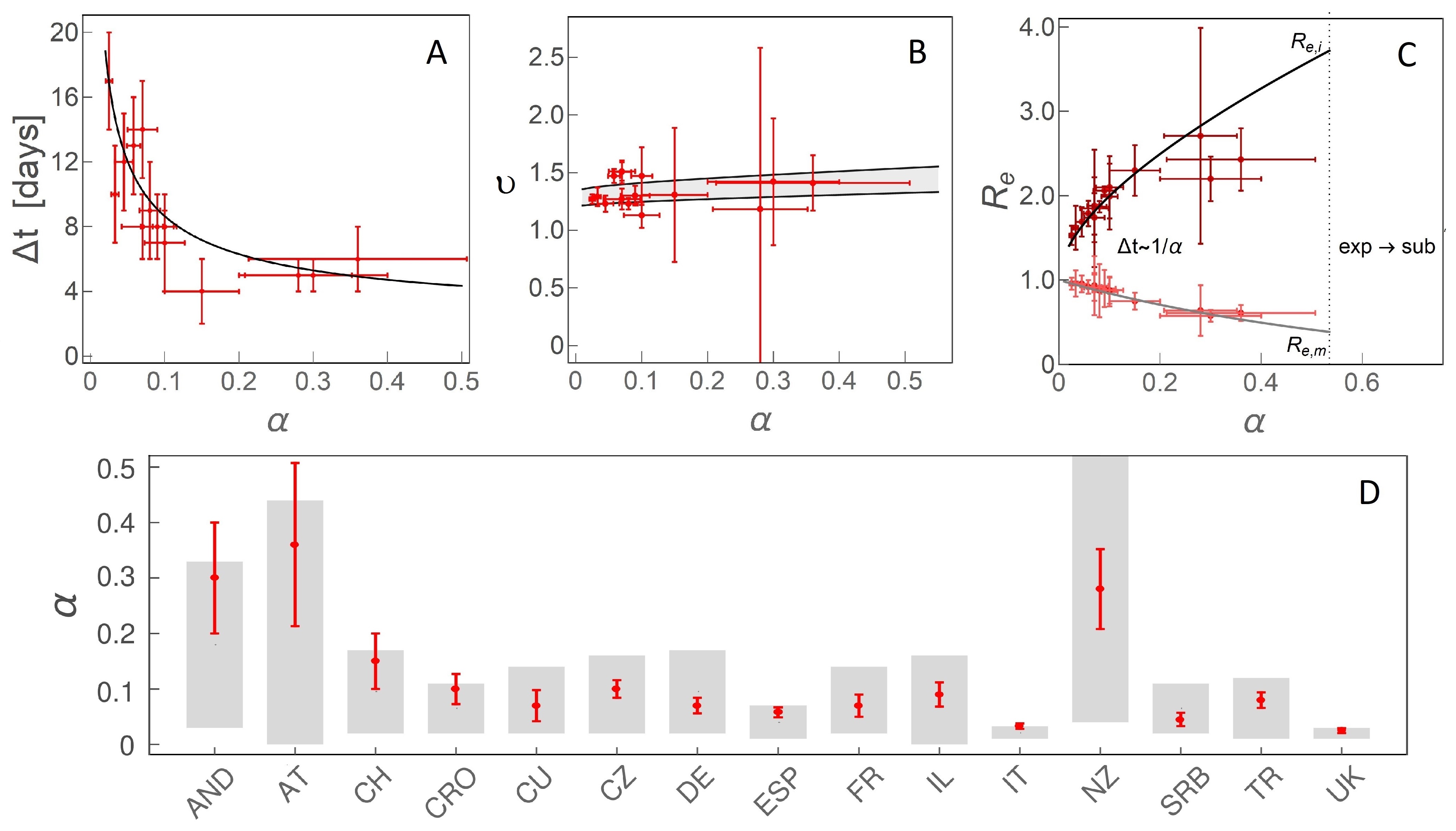}
  \caption{The dependence on the effective social distancing strength ($\alpha$) of A) $\Delta t$, the duration of the superlinear regime, B) $\upsilon$, the scaling exponent of the superlinear regime, C) $R_e$, effective reproduction number at the left inflection point ($R_{e,i}$) and the maximum ($R_{e,m}$) of $I(t)$. $\Delta t \sim 1/ \alpha$ indicates that the time, in which the change from $R_{e,i}$ to $R_{e,m}$ is exhibited, is approximately inversely proportional to $\alpha$. "$exp \rightarrow sub$" indicates the region of $\alpha$ where we predict a direct transition from exponential to sublinear growth. D) Comparison of $\alpha$ constrained from analytical derivations (the grey bands) and numerical analysis, with countries indicated on the horizontal axis by their abbreviations. Results obtained by independent numerical analysis are presented by red dots with  corresponding errorbars. }
  \label{fig:sl2}
\end{figure*}
How the effective reproduction number $R_e$ changes during this regime, i.e., between the left inflection point and the maximum of $I(t)$? $R_e$ quantifies the average number of secondary cases per infectious case, so that $R_e > 1$ signifies disease outburst, while for $R_e <1$ the disease starts to be eliminated from the population\textsuperscript{\cite{17}}. The Eq.~(\ref{EFT}) provides expressions for $R_{e,i}$ (at the inflection point) and $R_{e,m}$ (at the maximum). Interestingly, from Figure~2C, we observe that $R_{e,i}$ and $R_{e,m}$ do not depend on $R_{0,\rm{free}}$ and are, respectively, significantly larger and smaller than~1, which shows that transition from infection outburst to extinguishing happens during the superlinear growth. Consequently, the steepness of $R_e$ change over the superlinear regime significantly increases (larger change over smaller time interval, see Figure~2C)  with the measure strength.

Finally, in the {\it sublinear} regime, in  a wide vicinity of $I(t)$ maximum (which marks the beginning of the sublinear growth) leading non-linear term of $D(t)$ is cubic ($\sim t^3$, with negative prefactor). This is consistent with the expansion of $I(t)$ around $t_m$, which has leading negative quadratic ($t^2$) dependence (see Eqs.~(3) and~(\ref{MP_T})). The ratio between the prefactors in $D(t)$ expansion is given by Eq.~(\ref{jednakost3}), from which we see that $\alpha$ can be directly constrained, as shown in Figure 2D. For the 10 countries with consistent tracking of $D, A$ and $F$, we independently numerically determined $\alpha$ and compared it with analytical results coming from Eq.~(\ref{jednakost3}), obtaining an excellent agreement between our derivations and numerical results. The obtained $\alpha$ values should be understood as an effective epidemic containment measure - i.e., estimating the true result of the introduced measures, which can be used to evaluate the practical effectiveness of the official policies.

To demonstrate how constraining $\alpha$ can aid numerical analysis in the cases when $A$ is not continuously tracked, we next analyze five additional countries listed in the Outline section, so that altogether our study covers majority of COVID-19 hotspots, which (at the time of this analysis) are close to saturation in confirmed counts. Furthermore, in the specific cases of UK and Italy, where we analytically obtained both very low and very constrained $\alpha$ ($0.01 < \alpha < 0.04$), we chose five times larger parameter span in $\alpha$ in the numerical analysis, to confirm that these low values are indeed preferred by the exhaustive numerical search. For example, the finally obtained $\alpha$ for Italy ($0.033 \pm 0.005)$ and UK ($0.025 \pm 0.005$), together with previously obtained agreements shown in Figures~2A-C, strongly confirm that the observed growth patterns provide invaluable information for successful analysis of the infection progression data.

To further illustrate this, the synergy of analytical derivations and numerical analysis presented above enables us to, directly from the publicly available data, infer key infection parameters necessary to assess epidemics risks (provided in the Supplement (Table~S1)). We estimate these parameters by the same model/analysis, for a number of diverse countries, allowing their direct comparison. In Figure 3, we show together Case Fatality Rate (CFR), Infected Fatality Rate (IFR) and infection Attack Rate (AR)\textsuperscript{\cite{17,20}}. CFR is the number of fatalities per {\it confirmed} cases. CFR can, in principle, be inferred directly from the data, but since different countries are in different phases of infection, we project forward the number of confirmed cases until a saturation is reached for each country, from which we calculate CFR. IFR (crucial parameter for assessing the risks for infection progression under different scenarios) is the number of fatalities per total number of {\it infected} cases, which is a genuine model estimate, due to the unknown total number of infected cases. AR (necessary for understanding the virus recurrence risk) is also determined from our model and provides an estimate of the fraction of the total population that got infected and possibly resistant.

\begin{figure*}
  \includegraphics[width=\linewidth]{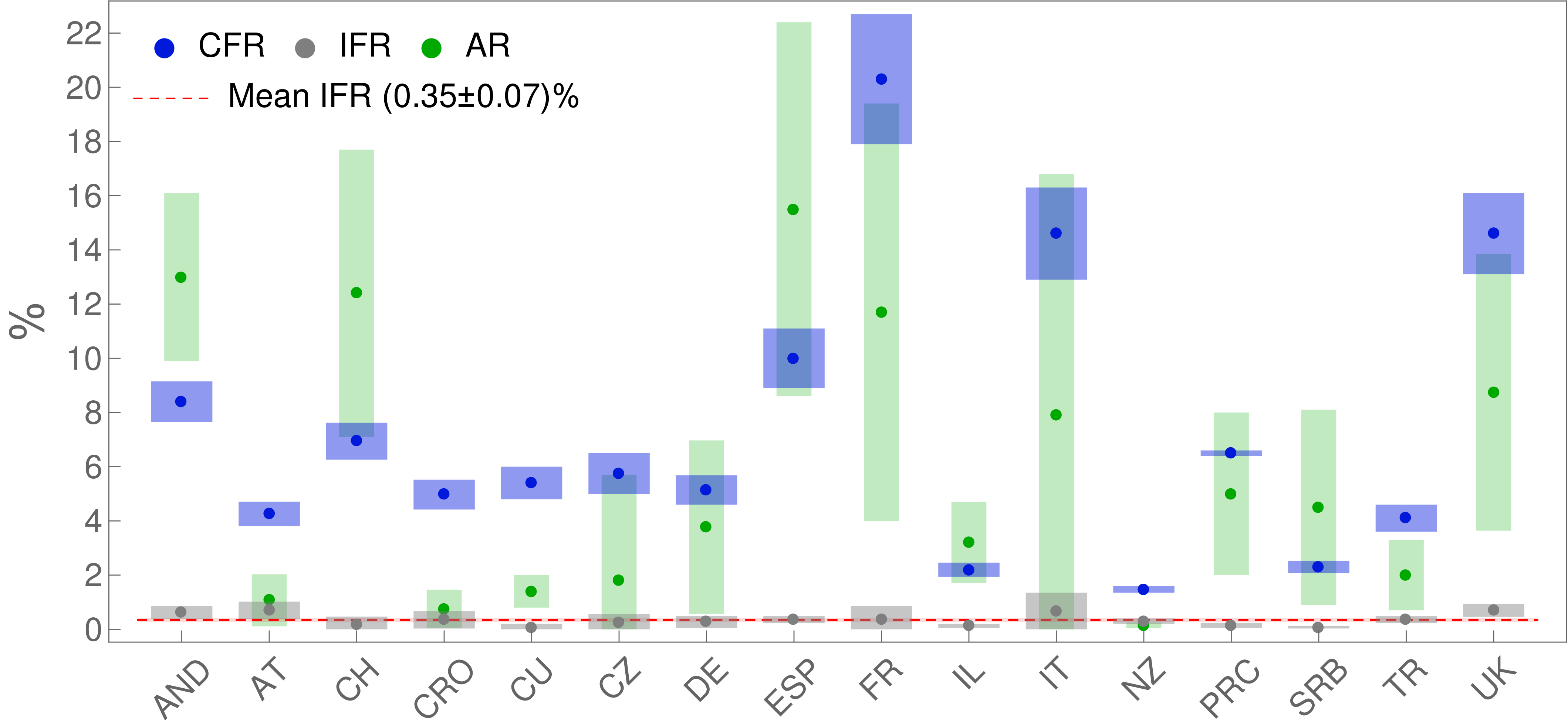}
  \caption{CFR, IFR and AR, inferred for countries whose abbreviations  are indicated on the horizontal axis, are denoted, respectively, by blue, grey and green dots, with errorbars indicated by corresponding bands. The dashed red horizontal line stands for IFR consistent with a mean value (indicated in the legend). Values for PRC are from\textsuperscript{\cite{21}}.}
  \label{fig:sl3}
\end{figure*}
From Figure 3, we see that CFR takes very different values for different countries, from below $2\%$ (New Zealand) to above $20\%$ (France). On the other hand, IFR is consistent with a constant value (the dashed red line in the figure) of $\sim 0.3-0.4\%$. In distinction to IFR, AR also takes diverse values for different countries, ranging from $\sim 1\%$ to as high as $\sim 15\%$ (though with large errorbars). Although diverse, these AR values are well bellow the classical herd immunity threshold of $60-70 \%$\textsuperscript{\cite{2,22}}, raising a major risk of infection recurrence with lifting social distancing measures. This conclusion is also consistent with independent experimental measurements, through serology tests, where low AR values are consistently obtained\textsuperscript{\cite{23}}; e.g. preliminary results of a large scale study in Serbia obtained AR of~6\% well consistent with our reported value ($5 \pm 4\%$). The fact that estimated AR and CFR take diverse values, while IFR is nearly constant, implies that the obtained IFR value may present an inherent COVID-19 property.

\textbf{\textit{Conclusion: }}
We here developed a novel quantitative framework through which we showed that: {\it i}) The emergence of three distinct growth regimes in COVID-19 case counts can be reproduced both analytically and numerically. {\it ii}) Typically, a brief superlinear regime is characterized by a sharp transition from outburst to extinguishing the infection, where effective reproduction number changes from much larger to much smaller than one; more effective measures lead to shorter superlinear growth, and to a steeper change of the effective reproduction number. {\it iii}) Scaling exponent of the superlinear regime is surprisingly uniform for countries with diverse environmental and demographic factors and epidemics containment policies; this highly non-trivial empirical result is well reproduced by our model. {\it iv}) Scaling prefactors in the sublinear regime contain crucial information
for analytically constraining infection progression parameters, so that they can be straightforwardly extracted through numerical analysis.  Interestingly, we found that the number of COVID-19 fatalities per total number of infected is highly uniform across diverse analyzed countries, in distinction to other (highly variable) infection parameters, and about twice higher than commonly quoted for influenza (0.3-0.4\% compared to 0.1-0.2\%), which may be valuable for direct assessment of the epidemics risks.

While state-of-the-art approach in epidemiological modeling uses computationally highly demanding numerical simulations, the results above demonstrate a shift of paradigm towards simpler, but analytically tractable models, that can both explain common dynamical features of the system and be used for straightforward and highly constrained parameter inference. This shift is based on a novel framework that relates universal growth patterns with characteristic points of the infective curve, followed by analytical derivations in the vicinity of these points, in an approach akin to those in a number of physics problems. The framework presented here can be, in principle, further extended towards e.g. including stochastic effects or different heterogeneities such as age-structure. However, these are non-trivial tasks, and it remains to be seen to what extent the analytical results can be obtained in those more complex models. Overall, as our approach does not depend on any COVID-19 specifics, the developed framework can also be readily applied to potential outbursts of future infections.

\textbf{\textit{Acknowledgements: }}
This work was supported by the Ministry of Education, Science and Technological Development of the Republic of Serbia.




\medskip

%

\begin{thebibliography}{9}
\bibitem{1} 
 WHO report. URL https://www.who.int/emergencies/diseases/novel-coronavirus-2019/situation.
 \bibitem{2} O. Diekmann, H. Heesterbeek, T. Britton, {\textit{Mathematical tools for understanding infectious disease dynamics}}, Princeton University Press, {\textbf{2012}}.
\bibitem{3} M. Martcheva, {\it An introduction to mathematical epidemiology} Springer, Berlin, Germany  {\bf 2015}.
\bibitem{4} M. J. Keeling, P. Rohani, {\it Modeling infectious diseases in humans and animals},  Princeton University Press, {\bf 2011}.
\bibitem{5} H. Tian, Y. Liu, Y. Li, C. H. Wu, B. Chen, U. G. Kraemer, B. Li, J. Cai, B. Xu, Q. Yang, B. Wang, P. Yang, Y. Cui, Y. Song, P. Zheng, Q. Wang, O. N. Bjornstad, R. Yang, B. T. Grenfell, O. G. Pybus, C. Dye, {\it Science} {\bf 2020},  {\it 368}, 638.
\bibitem{6} G. Chowell, L. Sattenspiel, S. Bansal, C. Viboud, {\it Physics of Life Reviews} {\bf 2016}, {\it 18}, 66.
\bibitem{Worldometer} Worldometer {\bf 2020}. {\it COVID-19 coronavirus pandemic} [Online]. Available at: https://www.worldometers.info/coronavirus/ [Accessed June 2, 2020].
\bibitem{7} M. Day, {\it BMJ: British Medical Journal (Online)} {\bf 2020}, {\it 368}, DOI: https://doi.org/10.1136/bmj.m1165.
\bibitem{8} G. N. Wong, Z. J. Weiner, A. V. Tkachenko, A. Elbanna, S. Maslov, N. Goldenfeld, {\it arXiv preprint} {\bf 2020}, {\it arXiv:200602036}.
\bibitem{9} J. R. Dormand, P. J. Prince, {\it Journal of computational and applied mathematics} {\bf 1980},  {\it 6}, 19.
\bibitem{24} J. M. Vilar, L. Saiz. {\textit{The evolving worldwide dynamic state of the COVID-19 outbreak.}} medRxiv {\bf 2020}.
\bibitem{25} N. C. Grassly, C. Fraser, {\it Nature Reviews Microbiology} {\bf 2008}, {\it 6}, 477.
\bibitem{10} R. Phillips, J. Kondev, J. Theriot, H. Garcia, {\it Physical biology of the cell}, Garland Science, New York, NY, USA {\bf 2012}.
\bibitem{11} F. Bowman,  {\it Introduction to Bessel Functions}, Dover Publications, New York, NY, USA {\bf 1958}.
\bibitem{12} D. Zwillinger, {\it Standard Mathematical Tables and Formulae},  CRC Press, Boca Raton, FL, USA {\bf 1995}.
\bibitem{13} M. Abramowitz, T. A. Stegun, {\it Handbook of Mathematical Functions}, (Ed: M. Abramowitz), Dover Publications, New York, NY, USA {\bf 1972}.
\bibitem{17} Y. M. Bar-On, A. I. Flamholz, R. Phillips, R. Milo, {\it eLife} {\bf 2020}, {\it 9}, e57309.
\bibitem{14} E. Dong, H. Du, L. Gardner, {\it The Lancet Infectious Diseases} {\bf 2020}, {\it 20}, 533.
\bibitem{15} R. Li, S. Pei, B. Chen, Y. Song, T. Zhang, W. Yang, J. Shaman, {\it Science} {\bf 2020} {\it 368}, 489.
\bibitem{16} X. He, E. H. Y. Lau, P. Wu, X. Deng, J. Wang, X. Hao, Y. C. Lau, J. Y. Wong, Y. Guan, X. Tan, X. Mo, Y. Chen, B. Liao, W. Chen, F. Hu, Q. Zhang, M. Zhong, Y. Wu, L. Zhao, F. Zhang, B. J. Cowling, F. Li, G. M. Leung, {\it Nature medicine} {\bf 2020},  {\it 26}, 672.
\bibitem{18} W. H. Press, B. P. Flannery, S. A. Teukolsky, W. T. Vetterling, {\it Numerical recipes: The art of scientific computing}, Cambridge University Press, Cambridge, UK {\bf 1986}.
\bibitem{19} R. W. Cunningham, {\it Computers in Physics} {\bf 1993}, {\it 7}, 570.
\bibitem{21} M. Djordjevic,  M. Djordjevic, I. Salom, A. Rodic, D. Zigic, O. Milicevic, B. Ilic, {\bf 2020},   {\it arXiv:2005.09630}.
\bibitem{20} S. Eubank, I. Eckstrand, B. Lewis, S. Venkatramanan, M. Marathe, C. L. Barrett, {\it Bulletin of Mathematical Biology} {\bf 2020}, {\it 82}, 1.
\bibitem{22} T. Britton, F. Ball, P. Trapman, {\bf 2020},  {\it arXiv:2005.03085}.
\bibitem{23} F. P. Havers, C. Reed, T. Lim, J. M. Montgomery, J. D. Klena, A. J. Hall, A. M. Fry, D. L. Cannon, C.-F. Chiang, A.Gibbons, I. Krapiunaya, M. Morales-Betoulle, K. Roguski, M. Ata Ur Rasheed, B. Freeman, S. Lester, L. Mills, D. S. Carroll, S. M. Owen, J. A. Johnson, V. Semenova, C. Blackmore, D. Blog, S. J. Chai, A. Dunn, J. Hand, S. Jain, S. Lindquist, R. Lynfield, S. Pritchard, T. Sokol, L. Sosa, G. Turabelidze, S. M. Watkins, J. Wiesman, R. W. Williams, S. Yendell, J. Schiffer, N. J. Thornburg, {\it JAMA Internal Medicine} {\bf 2020},  DOI: 10.1001/jamainternmed.2020.4130.
\end{thebibliography}

\end{document}